 \documentclass[journal]{IEEEtran}
\usepackage{epsfig}
\usepackage{graphicx}
\usepackage{amsmath}
\usepackage{amssymb}
\usepackage{amsmath} 
\usepackage{amsmath} 
\usepackage{algorithmic}
\usepackage{algorithm}
\usepackage{subfigure}
\usepackage{multirow}
\usepackage{url}
\usepackage{color}
\usepackage{hyperref}
\usepackage{balance}
\usepackage{setspace}
\usepackage{slashbox}
\usepackage{booktabs}
\usepackage{threeparttable}
\usepackage{stfloats}
\usepackage{textcomp}
\usepackage{pifont}
\usepackage{float}
\usepackage{array}
\renewcommand{\baselinestretch}{1.0}
\newcommand{\re}[1]{\textcolor{blue}{#1}}

\newcommand{\etal}{{\em et al.}}       
\newcommand{\eg}{{\em e.g.}}           
\newcommand{\ie}{{\em i.e.}}           
\newcommand{\etc}{{\em etc.}}         

\ifCLASSOPTIONcompsoc
  \usepackage[nocompress]{cite}
\else
  \usepackage{cite}
\fi

\ifCLASSINFOpdf
\else
\fi

\begin{document}
%
\title{Crosslink-Net: Double-branch Encoder Segmentation Network via Fusing Vertical and Horizontal Convolutions}
%
%
%
%

\author{Qian~Yu, Lei~Qi, Luping~Zhou, Lei~Wang, Yilong~Yin, Yinghuan~Shi, Wuzhang~Wang, Yang~Gao
\IEEEcompsocitemizethanks{
\IEEEcompsocthanksitem \emph{Corresponding Authors: Yinghuan Shi and Wuzhang Wang}.
\IEEEcompsocthanksitem Qian Yu is with School of Data and Computer Science, Shandong Women's University, Jinan 250300, China, and the State Key Laboratory for Novel Software Technology, Nanjing University, Nanjing 210023, China (e-mail: yuqian@sdwu.edu.cn).)
\IEEEcompsocthanksitem Lei~Qi is with School of Computer Science and Engineering, Southeast University, Nanjing 211189, China (e-mail: qilei@seu.edu.cn).
\IEEEcompsocthanksitem Yinghuan Shi and Yang Gao are with the State Key Laboratory for Novel Software Technology, Nanjing University, Nanjing 210023, China (e-mail: syh@nju.edu.cn; gaoy@nju.edu.cn).
\IEEEcompsocthanksitem Luping Zhou is with the School of Electrical and Information Engineering, University of Sydney, NSW 2006, Australia (e-mail: luping.zhou@sydney.edu.au).
\IEEEcompsocthanksitem Lei~Wang is with the School of Computing and Information Technology, University of Wollongong, Wollongong, NSW 2522, Australia (e-mail: leiw@uow.edu.au).
\IEEEcompsocthanksitem Yilong~Yin is with the School of Soft, Shandong University, Jinan 250100, China(e-mail: ylyin@sdu.edu.cn).
\IEEEcompsocthanksitem Wuzhang~Wang is with Department of Respiratory and Critical Care Medicine, Shandong Chest Hospital, Jinan 250013, China (e-mail: wz.wang.sd@gmail.com).
}
}




\IEEEtitleabstractindextext{%
\begin{abstract}
Accurate image segmentation plays a crucial role in medical image analysis, yet it faces great challenges of various shapes, diverse sizes, and blurry boundaries. To address these difficulties, square kernel-based encoder-decoder architecture has been proposed and widely used, but its performance remains still unsatisfactory. To further cope with these challenges, we present a novel double-branch encoder architecture. Our architecture is inspired by two observations: 1) Since the discrimination of features learned via square convolutional kernels needs to be further improved, we propose to utilize non-square \textit{vertical} and \textit{horizontal} convolutional kernels in the double-branch encoder, so features learned by the two branches can be expected to complement each other. 2) Considering that spatial attention can help models to better focus on the target region in a large-sized image, we develop an attention loss to further emphasize the segmentation on small-sized targets. Together, the above two schemes give rise to a novel double-branch encoder segmentation framework for medical image segmentation, namely Crosslink-Net. The experiments validate the effectiveness of our model on four datasets. The code is released at \textcolor{blue}{\url{https://github.com/Qianyu1226/Crosslink-Net}}.

\end{abstract}

}

\maketitle
\IEEEdisplaynontitleabstractindextext
\IEEEpeerreviewmaketitle
\bigskip
\bigskip

\IEEEraisesectionheading{\section{Introduction}}
\label{sec:introduction}

\begin{figure}[ht]
 \centering
 \setlength{\abovecaptionskip}{-0.05cm}
  \includegraphics[width=0.47\textwidth]{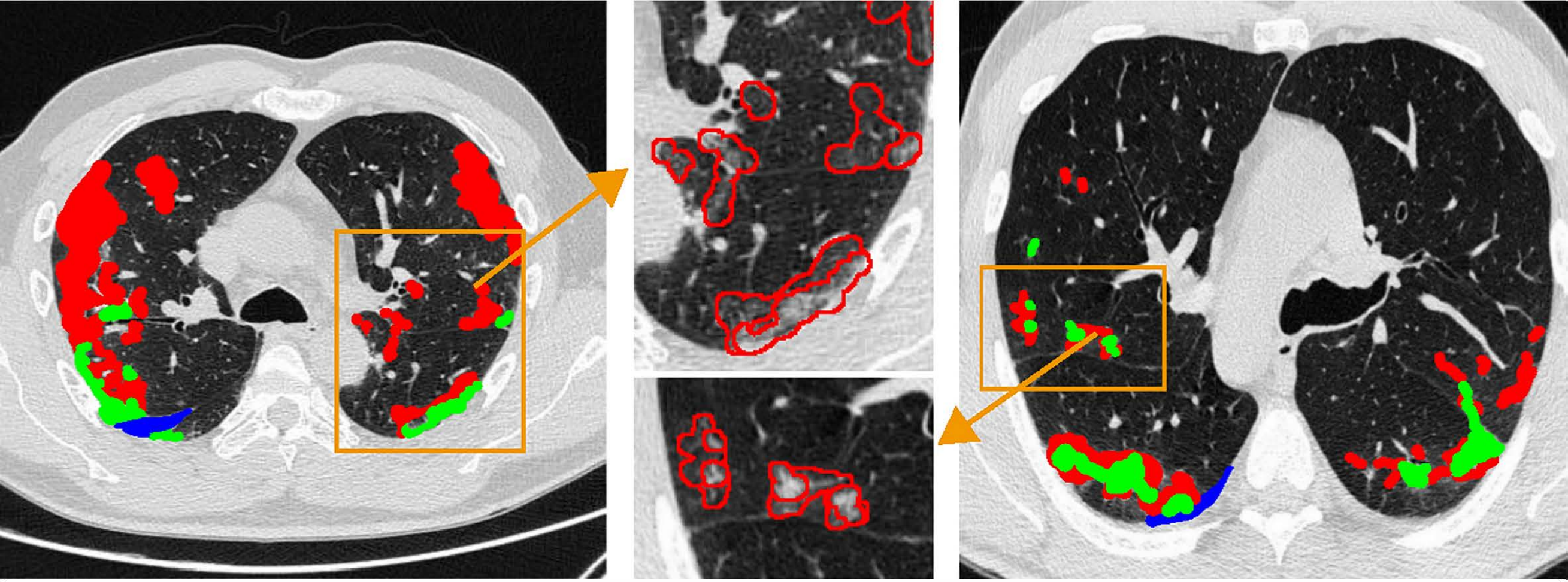}
 \caption{Examples of COVID-19 CT images. The red, green, and blue regions are ground truth of ground-glass, consolidation and pleural effusion, respectively.}
 \label{fig:exm}
\vspace{-0.3cm}
\end{figure}

\IEEEPARstart{A}{iming} to provide position, size and appearance information to physician for subsequent treatment, accurate automatic medical image segmentation plays a critical role and is in great demand in clinical application. It is known that, a promising automatic segmentation model can provide stable segmentation results and largely reduce the time and cost of manual delineation \cite{shi2017does}. In recent years, deep learning-based methods have been widely used in medical image segmentation, while models based on the \emph{encoder-decoder skip-connection-linked architecture} (encoder-decoder models will be used for simplicity in the following part) \cite{zhou2020high, ronneberger2015u, gu2019ce-net:, yu2020mixmodule} are the most popular ones, particularly, the U-Net \cite{ronneberger2015u} and its variants  \cite{milletari2016v,Huang20193d,perslev2019one}.

Despite of the considerable improvement made by these encoder-decoder models, there are still several issues that still hinder their deployment in the real-life scenarios:

1) The existing encoder-decoder models are not always robust in capturing various shapes and size information, especially when it comes to targets with blurry boundaries. Zhou \etal \cite{zhou2020high} have pointed out that this is probably because the downsampling operation usually causes the loss of location details. Another reason could be that the low-level representation in shallow layers directly passed to the corresponding layers in decoder by skip connections might deteriorate the performance of the encoder-decoder model.

2) As discussed in \cite{zhou2020high, Murugesan2020a}, the popularly-used cross entropy loss used as major loss function in encoder-decoder models is sensitive to foreground-background imbalance\cite{Murugesan2020a}, so U-Net and its variants tend to easily neglect small-sized target regions.

Generally, in medical image segmentation tasks, the same target among different subjects usually shows large variations of shapes and sizes. Taking the COVID-19 CT image segmentation in Fig. \ref{fig:exm} as an example, the lung infections in different subjects show considerable differences in appearances. Furthermore, most lung infections have not only weak or even vanishing boundaries but also isolated small parts. Similar characteristics are also observed in other segmentation tasks, such as kidney tumors, pancreas, \etc.

 To address these issues, Zhou \etal \cite{zhou2020high} propose a multi-scale dense connections model with dilated convolution \cite{yu2017dilated} to highlight the boundary and isolated small target segmentation. In their model, besides contracting and expanding patches and their corresponding skip connections, four new eleven-convolution-layer-network skip connections are also embedded. Although it might be time consuming to fully train such a sophisticated model, their model proves that dilated convolution is effective in improving the performance of encoder-decoder architectures. Meanwhile, Gu \etal \cite{gu2019ce-net:} employ dense dilated convolutions to improve the performance of a context encoder network (CE-Net). We notice that, the dilated convolution used in \cite{gu2019ce-net:,zhou2020high} is to model the long-range context of targets \cite{hou2020strip}. Similarly, Yu \etal \cite{yu2020mixmodule} design multi-scale convolution kernels upon encoder-decoder structure to capture information from different levels. Unfortunately, both the multi-scale and dilated convolution operations utilize square kernels, largely limiting their ability to capture the anisotropy context \cite{hou2020strip}. However, the anisotropy context is very important for medical image segmentation, especially for nonrigid targets with various orientations and different aspect ratios \cite{man2019deep}, \eg, pancreas, kidney tumors, and the aforementioned lung infections. Yet, the small square kernel would neglect sufficient context, while a large one might bring in target-irrelevant information. Therefore, how to design a promising convolution kernel to tackle these issues becomes a critical yet challenging task.

\begin{figure}
 \centering
 \setlength{\abovecaptionskip}{-0.2cm}
 \includegraphics[width=0.44\textwidth]{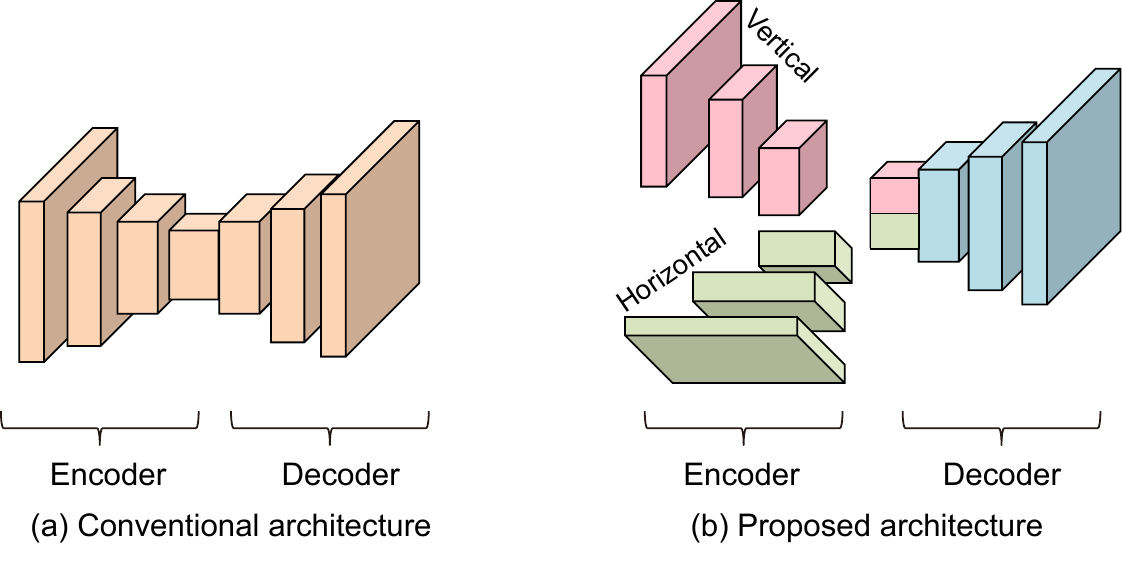}
 \caption{Comparison between different frameworks. Note that the proposed framework has double encoders with both vertical and horizontal information.}
 \label{fig:encoder}
\vspace{-0.3cm}
\end{figure}

In this paper, we propose a novel encoder-decoder model with our newly developed convolutional kernels and loss function, namely \textbf{Crosslink-Net}, to capture robust features of targets with various shapes and sizes. As shown in Fig. \ref{fig:encoder}, different from conventional encoder-decoder architecture that employs single-pathway encoder, we adopt two encoders to extract both vertical and horizontal information from cross-direction, as motivated by the following two observations.

Firstly, considering that shape information could not be effectively captured from typical square kernel popularly adopted in conventional works, we design a novel non-square kernel, termed as \emph{crosslink-kernel}, to contain a series of vertical and horizontal kernels. Our proposed crosslink-kernel can obtain relatively large receptive fields along the long side meanwhile capture local detail information along the short side. Thus, not only can the global shape information be provided by modeling long-range dependency from cross-direction, target-irrelevant information can also be prevented simultaneously. In addition, detailed (local) and contextual (global) information captured by crosslink-kernel is also vital for blurry boundary segmentation\cite{zhou2020high}.

To further exploit these kernels and observe the target shape from distinct perspective, it is necessary to adopt a double-branch encoder structure with one branch containing horizontal kernels while the other vertical kernels. This motivation can guarantee that: 1) each branch can concentrate on learning features from its own (vertical or horizontal) direction; 2) complementary information can be explored to fully describe various shapes. Whereas, using these kernels in a single-pathway encoder will result in a ``fat'' (wide) network and disordered learned information, which is not beneficial for segmentation, as Section \ref{sub:eva} will validate.

Secondly, regarding that spatial attention can emphasize small target regions in a large-size image \cite{oktay2018attention}, a new attention loss is investigated according to our double-branch encoder-decoder structure to enable the model to focus more on the location and shape of the small target. The loss is expected to ensure that the high-score information from both branches is as closely consistent to the ground truth as possible, so as to further improve the performance of the entire network.

The major contributions of our work are as follows:
\begin{enumerate}
\item We propose a double-branch encoder segmentation network, to better treat the variation of shapes with blurry boundary from cross-direction via non-square vertical and horizontal kernels.
\item We develop an attention loss to better touch the small target regions, in order to further enhance the discrimination of features and make the model robust to small targets.
\item Extensive experiments have verified the effectiveness of our model. Appealing results are obtained in lung infection, kidney tumor and left ventricle (LV) segmentation tasks.
\end{enumerate}

Our paper is organized as follows. In Section \ref{sec:related}, we briefly review related works on encoder-decoder models and non-square kernels. In Section \ref{sec:method}, we describe the technical details of the proposed Crosslink-Net. In Section \ref{sec:experiments}, we discuss the ablation analyses and experimental comparisons. Finally, we draw a conclusion in Section \ref{sec:conclusion}.

\section{Related Work}
\label{sec:related}
Since our model is based on the encoder-decoder architecture with non-square residual kernels being employed, we will review the progress on encoder-decoder architecture and non square kernels in recent years.
\subsection{Encoder-decoder Architecture}
The encoder-decoder architecture has demonstrated its effectiveness in medical image segmentation. Conventional encoder-decoder methods typically consist of a contracting path and a symmetric expanding path with the skip connections. According to whether the contracting path is pretrained or not, current encoder-decoder methods can be roughly divided into two categories: the U-Net-like model and the backbone-FPN model.

\textbf{U-Net-like model:} As one of the representative encoder-decoder models, U-Net \cite{ronneberger2015u} has been widely used in various medical image segmentation tasks. A lot of medical image segmentation models are either its variants or inspired by it. For example, V-Net \cite{milletari2016v} is the 3D form of U-Net by performing volumetric convolutions, which performs well in prostate segmentation. Subsequently, 3D-U$^{2}$-Net \cite{Huang20193d} is developed from V-Net by embedding a domain adapter to each block of encoder and decoder. MPUnet \cite{perslev2019one}, as an ensemble model, trains several U-Nets on multi-planar and obtains a high accuracy in different 3D segmentation tasks. In addition, Lalonde \etal \cite{lalonde2018capsules} also design a model based on U-Net with capsules, which achieves promising results in several segmentation tasks. H-DenseUNet \cite{li2018h-denseunet:} is a hybrid U-Net model fusing 2D and 3D features for liver tumor segmentation in CT images. Dong \etal \cite{dong2020texnet} propose a TexNet on U-Net architecture with attention blocks on decoder to segment gastric antrum ultrasound images. Similarly, Yu \etal \cite{yu2020c2fnas} present a U-Net-like neural architecture search to overcome 3D segmentation tasks in a coarse-to-fine manner.

\textbf{Backbone-FPN model:} This type of model generally employs some pre-trained networks as their backbones and connects one (or several) additional feature map pyramid networks (FPN) \cite{lin2017feature} at the tail. Typically, they first adopt the pre-trained classical model (\eg, VGG16 \cite{simonyan2014very}, ResNet-101 and ResNet-34 \cite{he2016deep}) as the backbone, and then use skip connections and different FPN to complete segmentation tasks. For instance, HD-Net~\cite{jia2019hd-net:} employs ResNet-101 and one FPN architecture to segment prostate MR images. SS-Net \cite{huo2019splenomegaly} segments abdominal MRI scans, with a ResNet-50 based encoder and a boundary refining decoder with a PatchGAN \cite{isola2017image} as the discriminator to supervise the training procedure. Ma \etal \cite {ma2019neural} design a cardiovascular MR image segmentation model with ResNet-101 and VGG-16 networks as its backbones. In these models, the pretrained backbones could reduce the demand for training data in some cases. However, as the popularly-used backbones are originally designed for natural image classification tasks, these models might require additional modules to eliminate inevitable domain shift between natural images and medical images.

 In these two types of architectures, the skip connections provide a global fusion of low-level and high-level features, which is crucial for the segmentation of targets \cite{zhou2020high}. Therefore, it becomes a type of model widely used in medical image segmentation. According to our best knowledge, the existing models adopt a single-pathway encoder and employ square kernels to learn features, without direction information being utilized. Compared with these models, the major improvement of our model lies in the fact that our encoder is double-branch, employing both vertical and horizontal kernels to learn complementary information from two directions.

\subsection{Non-square Kernels}
Recently, non-square kernels have triggered great interests among scholars. Szegedy \etal \cite{szegedy2016rethinking} adopt $n \times 1$ kernel connected with $1 \times n$ kernel to accomplish $n \times n$ convolution in the Inception V2, since then, many scholars utilize this inspiration to develop their new models. For instance, Huang \etal \cite{huang2019ccnet:} propose a criss-cross attention module to aggregate long-range pixel-wise contextual information in the horizontal and vertical directions. However, huge memory is required for computing the large affinity matrix at each spatial position \cite{hou2020strip}. To tackle this problem, Hou \etal \cite{hou2020strip} propose a strip pooling with $n \times 1$ and $1 \times n$ kernels to segment banded targets. Mou \etal \cite{mou2019cs} explore a CS-Net that implements a $3 \times 1$ and $1 \times 3$ convolution to capture boundary information of blood vessels from both horizontal and vertical directions. The problem with strip pooling \cite{hou2020strip} and CS-Net \cite{mou2019cs} is that they utilize non-square kernels merely as the attention factors in a form of a stand-alone module, hindering the full use of the non-square kernels in feature learning. In addition, the long-ranged information captured by strip pooling kernels in \cite{hou2020strip} includes the whole row or column context of an image, with the possibility of neglecting the small target, hence, cannot be directly applied to medical images.

Different from the above non-square kernel models \cite{huang2019ccnet:, hou2020strip, mou2019cs}, we employ non-square kernels directly in our encoder to learn features, rather than process features with extra non-square kernel attention modules. We have adopted non-square kernels together with crossbar and crossover patches in our previous works \cite{yu2019crossbar, yu2020crossover} which are non-encoder-decoder models that have achieved desirable performances in several segmentation tasks. In this work, our crosslink-kernel continues to experimentally show advantages over the square kernels and the non-square-kernel attentions.

\section{Method}
\label{sec:method}
\begin{figure*}[ht]
 \centering
  \includegraphics[width=0.98\textwidth]{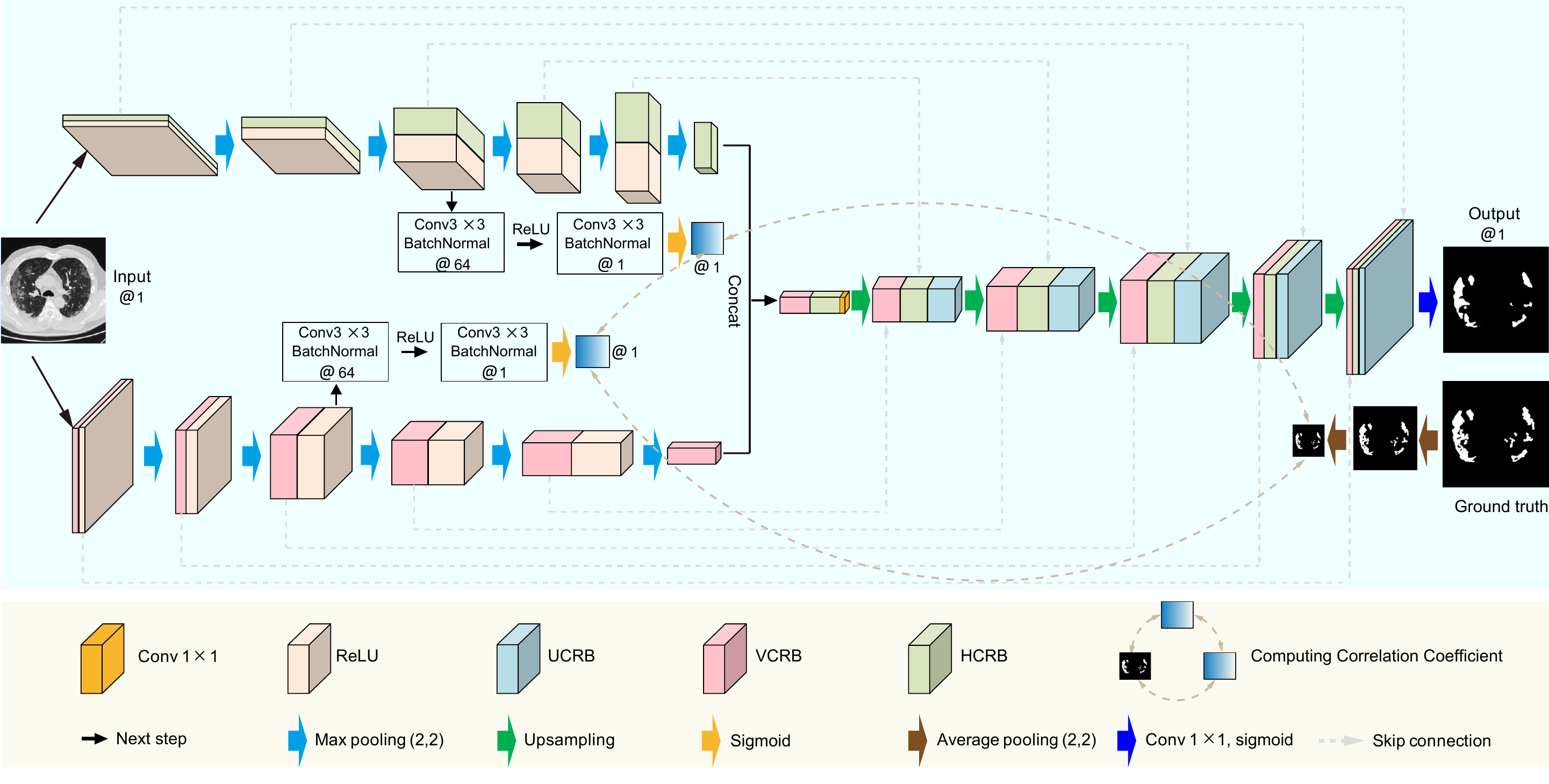}
 \caption{Framework of Crosslink-Net. The VCRB, HCRB, UCRB are vertical, horizontal, and up convolution residual blocks, respectively.}
 \label{fig:framework}
\vspace{-0.3cm}
\end{figure*}

As shown in Fig. \ref{fig:framework}, the architecture of our proposed Crosslink-Net consists of vertical and horizontal branches, with each branch containing five convolutional residual blocks. The blocks in the vertical branch mainly make up of vertical kernels, termed as \emph{Vertical Convolutional Residual Blocks} (\textbf{VCRB}). Similarly, the corresponding horizontal blocks are termed as \emph{Horizontal Convolutional Residual Blocks} (\textbf{HCRB}). Accordingly, the blocks in each branch are named as VCRB1, VCRB2, ..., VCRB5 and HCRB1, HCRB2, ..., HCRB5, respectively. The decoder also includes five blocks termed as \emph{Up Convolution Residual Blocks} (named as UCRB1, UCRB2, ..., UCRB5, respectively).

Specifically, as shown in Fig. \ref{fig:framework}, in Crosslink-Net, both branches have the same image as their input. In both branches, each convolution residual block is followed with a maximum pooling layer after the ReLU operation, so as to simultaneously reduce the dimension of its feature map, increase its receptive field, and strengthen its non-linearity. The pooling kernel is $2 \times 2$ with stride 2. The feature maps of VCRB3 and HCRB3 are used to calculate the loss function after the ReLU operation. In our decoder, a bilinear interpolation is utilized in the up-sampling operation.

\subsection{The Proposed Architecture}
\label{subsec:CRB}
The crosslink-kernels are designed for two purposes: to handle the variant appearance from cross-directions and to simultaneously learn local and contextual information for blurry and vanished boundaries. Hence, the $5 \times 3$ and $3 \times 1$ kernels are used to encode receptive field information of two scales in vertical direction, with the $3 \times 5$ and $1 \times 3$ kernels as their counterparts in the horizontal branch. Furthermore, to obtain abundant receptive fields and fine boundary information, we also adopt $3 \times 3$ kernels in both branches.

\begin{figure*}[ht]
 \centering
 \setlength{\abovecaptionskip}{-0.2cm}
  \includegraphics[width=0.75\textwidth]{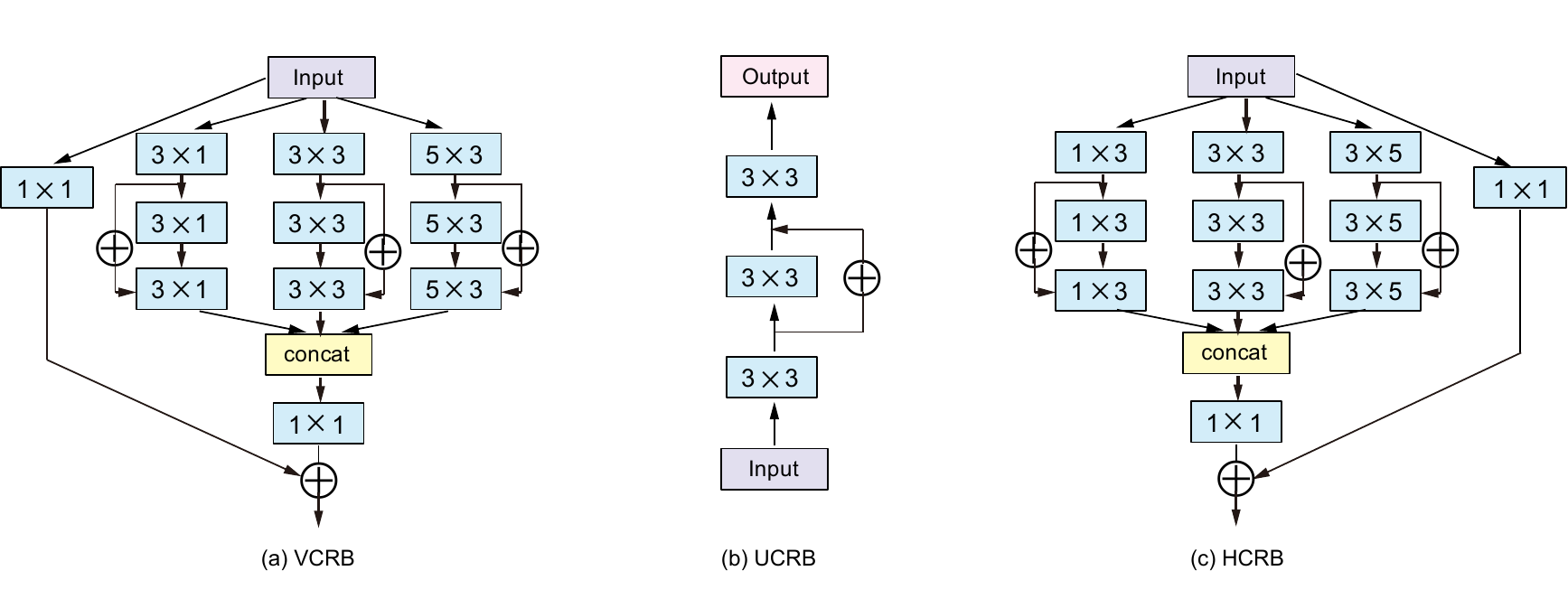}
 \caption{The vertical, up, and horizontal residual convolution blocks in Crosslink-Net. $\bigoplus $ means element-wise addition.}
 \label{fig:blocks}
\vspace{-0.3cm}
\end{figure*}

In the encoder, most convolution operations are performed three times consecutively to obtain deeper information. To prevent the loss of boundary information and speed up convergence, residual connections and $1 \times 1$ kernels are employed in each convolutional block. The specific structure of each block in the encoder is shown in Fig. \ref{fig:blocks} (a) and (c). In the decoder, each basic block is composed of three continuous convolutions with $3 \times 3$ kernels and residual connection, the structure of which is shown in Fig. \ref{fig:blocks} (b). Each convolution operation is followed by Batch Normalization and ReLU operation, except for the $1 \times 1$ convolution in front of $\oplus$.

In Table \ref{tab:blocks}, we provide the parameter settings of VCRB1, VCRB2, $\cdots$, VCRB5, HCRB1 and all UpCRBs. The Conv2d denotes a single convolution operation. The RConv2d indicates three consecutive convolution operations with a residual connection. The parameters in parentheses of Conv2d and RConv2d are the input channels, output channels, kernel size, padding value, and states of ReLU. The stride and BatchNorm are set as 1 and true, respectively. The parameters of the remaining HCRB are set with the same values as their corresponding VCRB counterparts, except for kernels.

\begin{table}
\caption{The parameters of each convolution residual block.}
\center
\label{tab:blocks}
\renewcommand\arraystretch{1.2}
\scriptsize{
\begin{tabular}{m{0.2in}<{\centering}|m{1.6in}<{\centering}|m{0.8in}<{\centering}|m{0.12in}<{\centering}}
\toprule
\multicolumn{1}{l|}{Blocks} & step1  & step2 & step3 \\
\midrule
\multirow{4}{*}{VCRB1}       & \multicolumn{2}{l|}{Conv2d(1,32, (1,1), (0,0), ReLU=False)}                                                                                                                                                                                         & \multirow{4}{*}{Add} \\ \cline{2-3}
                             & \begin{tabular}[c]{@{}l@{}}RConv2d(1,32,(3,1),(1,0),ReLU=True)\\ RConv2d(1,32,(3,3),(1,1),ReLU=True)\\ RConv2d(1,32,(5,3),(2,1),ReLU=True)\end{tabular} & \begin{tabular}[c]{@{}l@{}}Concat,\\ Conv2d(96,32,(1,1),\\ (0,0),ReLU=False)\end{tabular} &\\
\midrule
\multirow{4}{*}{HCRB1}       & \multicolumn{2}{l|}{Conv2d(1,32, (1,1), (0,0), ReLU=False)}                                                                                                                                                                                         & \multirow{4}{*}{Add} \\ \cline{2-3}
                             & \begin{tabular}[c]{@{}l@{}}RConv2d(1,32,(1,3),(0,1),ReLU=True)\\ RConv2d(1,32,(3,3),(1,1),ReLU=True)\\ RConv2d(1,32,(3,5),(1,2),ReLU=True)\end{tabular} & \begin{tabular}[c]{@{}l@{}}Concat,\\ Conv2d(96,32,(1,1),\\ (0,0),ReLU=False)\end{tabular} &\\
\midrule
\multirow{4}{*}{VCRB2}       & \multicolumn{2}{l|}{Conv2d(32,64, (1,1), (0,0), ReLU=False)}                                                                                                                                                                                         & \multirow{4}{*}{Add} \\ \cline{2-3}
                             & \begin{tabular}[c]{@{}l@{}}RConv2d(32,64,(3,1),(1,0),ReLU=True)\\
                             RConv2d(32,64,(3,3),(1,1),ReLU=True)\\ RConv2d(32,64,(5,3),(2,1),ReLU=True)\end{tabular} & \begin{tabular}[c]{@{}l@{}}Concat,\\ Conv2d(192,64,(1,1),\\ (0,0),ReLU=False)\end{tabular} &\\
\midrule
\multirow{4}{*}{VCRB3}       & \multicolumn{2}{l|}{Conv2d(64,128, (1,1), (0,0), ReLU=False)}                                                                                                                                                                                         & \multirow{4}{*}{Add} \\ \cline{2-3}
                             & \begin{tabular}[c]{@{}l@{}}RConv2d(64,128,(3,1),(1,0),ReLU=True)\\
                             RConv2d(64,128,(3,3),(1,1),ReLU=True)\\ RConv2d(64,128,(5,3),(2,1),ReLU=True)\end{tabular} & \begin{tabular}[c]{@{}l@{}}Concat,\\ Conv2d(384,64,(1,1),\\ (0,0),ReLU=False)\end{tabular} &\\
\midrule
\multirow{4}{*}{VCRB4}       & \multicolumn{2}{l|}{Conv2d(128,256,(1,1), (0,0), ReLU=False)}                                                                                                                                                                                         & \multirow{4}{*}{Add} \\ \cline{2-3}
                             & \begin{tabular}[c]{@{}l@{}}RConv2d(128,256,(3,1),(1,0),ReLU=True)\\
                             RConv2d(128,256,(3,3),(1,1),ReLU=True)\\ RConv2d(128,256,(5,3),(2,1),ReLU=True)\end{tabular} & \begin{tabular}[c]{@{}l@{}}Concat,\\ Conv2d(768,64,(1,1),\\ (0,0),ReLU=False)\end{tabular} &\\
\midrule
\multirow{4}{*}{VCRB5}       & \multicolumn{2}{l|}{Conv2d(256,512,(1,1), (0,0), ReLU=False)}                                                                                                                                                                                         & \multirow{4}{*}{Add} \\ \cline{2-3}
                             & \begin{tabular}[c]{@{}l@{}}RConv2d(256,512,(3,1),(1,0),ReLU=True)\\
                             RConv2d(256,512,(3,3),(1,1),ReLU=True)\\ RConv2d(256,512,(5,3),(2,1),ReLU=True)\end{tabular} & \begin{tabular}[c]{@{}l@{}}Concat,\\ Conv2d(1536,64,(1,1),\\ (0,0),ReLU=False)\end{tabular} &\\
\midrule
UCRB5&\multicolumn{3}{c}{RConv2d($512 \times 3$, 256, (1,1), (0,0), ReLU=True)}\\
\midrule
UCRB4&\multicolumn{3}{c}{RConv2d($256 \times 3$, 128, (1,1), (0,0), ReLU=True)}\\
\midrule
UCRB3&\multicolumn{3}{c}{RConv2d($128 \times 3$, 64, (1,1), (0,0), ReLU=True)}\\
\midrule
UCRB2&\multicolumn{3}{c}{RConv2d ($64 \times 3$, 32, (1,1), (0,0), ReLU=True)}\\
\midrule
UCRB1&\multicolumn{3}{c}{RConv2d ($32 \times 3$, 16, (1,1), (0,0), ReLU=True)}\\
\bottomrule
\end{tabular}
}
\end{table}

\begin{figure*}
 \centering
 \setlength{\abovecaptionskip}{-0.05cm}
  \includegraphics[width=0.97\textwidth]{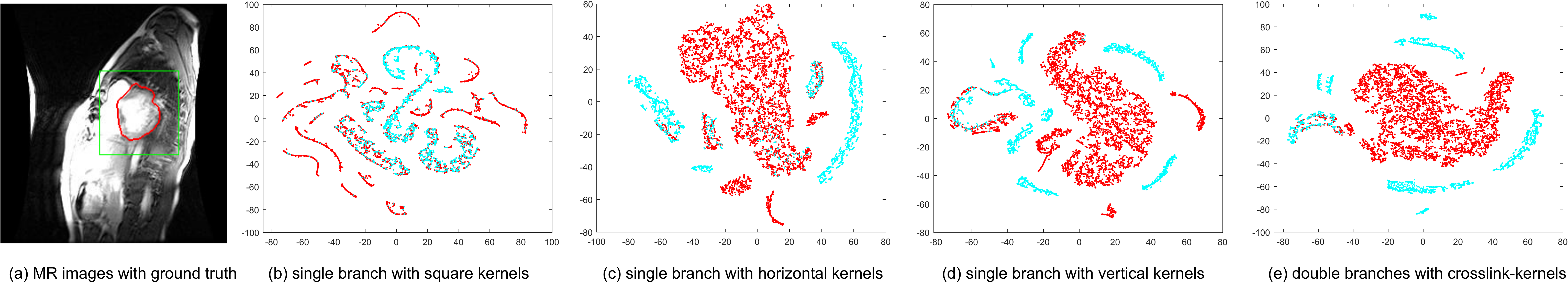}
 \caption{The t-SNE results of different models. The red and cyan points are background and target pixels in the green box of each MR image.}
 \label{fig:tsne}
\vspace{-0.3cm}
\end{figure*}

To highlight the advantages of non-square kernels and double-branch encoder, we compare the predictions of models with square kernel single branch, horizontal kernel single branch, vertical kernel single branch and crosslink-kernel double-branch. All parameters are set according to Table \ref{tab:blocks} except for kernels. The square kernel is the most conventional $3 \time 3$ kernel. The t-SNE (t-distributed Stochastic Neighbour Embedding) \cite{maaten2008visualizing} is adopted to evaluate these predictions. Fig.\ref{fig:tsne} illustrates the t-SNE of pixels in green box of three MR images. As shown, the background and target cases predicted by square kernel model are almost indivisible in Fig. \ref{fig:tsne} (b), while the cases in Fig. \ref{fig:tsne} (c) and Fig. \ref{fig:tsne} (d) are much desirable, indicating the effectiveness of non-square kernels. Fig. \ref{fig:tsne} (e) demonstrates the necessity of integrating the vertical and horizontal branches.

\subsection{Attention Loss}
\label{sub:loss}
In our method, considering that vertical and horizontal branches have the same image as their input, although learning features from different directions, the learning results of them should be consistent with the corresponding ground truth. To further enhance this consistency, we require the features learned by VCRB3 to be similar to those learned by HCRB3. The features learned from these middle blocks are utilized to extract accurate location and discriminative information, since the shallow blocks might learn more noisy information, while the deep blocks focus on learning high-level features. Compared with these two types of blocks, the middle block might capture quite favorable information about the shape and position of the target. We achieve the above goal through two steps: 1) calculating the spatial attention map of VCRB3 and HCRB3, and 2) forcing the two maps to correlate to the ground truth, \ie, maximizing their correlation coefficient.

The spatial attention map of a certain layer refers to a single-channel image composed of attention coefficients (positive value between 0 and 1). The resolution of the attention map is the same as that of the single-feature map of the same layer. The attention coefficient indicates the importance score of each point in the feature map. Thus, the target could be emphasized by the model, ideal for segmenting small targets. The schematic diagram of the attention map has been illustrated in Fig. \ref{fig:framework}.

Wang \etal \cite{wang2019automatic} and Xie \etal \cite{xie2015holistically-nested} have designed their loss functions by minimizing the difference between the attention map and the manual annotation. The former \cite{wang2019automatic} computes maps on each scale while the latter \cite{xie2015holistically-nested} only processes attention map on the last layer. Inspired by these two methods, our proposed attention loss aims to maximize the correlation coefficient between the attention map of VCRB3, HCRB3 and the ground truth. Our purpose is to strengthen the correlation among information learned by vertical, horizontal branches and the ground truth. These middle-level blocks could abandon partial low-level features while preserving the shape and position information. We will verify the advantage of the middle level in Section \ref{sub:eva}.

Specifically, let $\mathbf{M}$ denote the mask image, which could be achieved by twice average pooling of the ground truth with $(2,2)$ pooling kernel and striding of 2. $\mathbf{A}_\text{V}$ and $\mathbf{A}_\text{H}$ denote the attention maps of VCRB3 and HCRB3, respectively. Note that, $\mathbf{M}$, $\mathbf{A}_\text{V}$, and $\mathbf{A}_\text{H}$ are with the same dimensionality. Given $H$ and $W$ denote the height and width of the feature map of VCRB3 and HCRB3, we define
\begin{displaymath}
\mathbf{T}_\text{V}=\texttt{vec}(\mathbf{A}_\text{V}-\mu_{\mathbf{A}_\text{V}}),
\end{displaymath}
\begin{displaymath}
\mathbf{T}_\text{H}=\texttt{vec}(\mathbf{A}_\text{H}-\mu_{\mathbf{A}_\text{H}}),
\end{displaymath}
\begin{displaymath}
\mathbf{T}_\text{M}=\texttt{vec}(\mathbf{M}-\mu_{\mathbf{M}}),
\end{displaymath}
where $\mu_{\mathbf{A}_\text{V}}$, $\mu_{\mathbf{A}_\text{H}}$, and $\mu_{\mathbf{M}}$ denote the mean value of $\mathbf{A}_\text{V}$, $\mathbf{A}_\text{H}$, and $\mathbf{M}$, respectively. The $\texttt{vec} (\cdot)$ indicates the vectorization operation of a matrix. Then, our proposed attention loss could be defined as:
\begin{equation}
\label{eq4}
{L}_\text{atn}=-\frac{(\mathbf{T}_\text{V} \times \mathbf{T}_\text{H} \times \mathbf{T}_\text{M}) \cdot \mathbf{1}}{\sqrt{(\mathbf{T}_\text{V} \cdot \mathbf{T}_\text{V}) (\mathbf{T}_\text{H} \cdot \mathbf{T}_\text{H}) (\mathbf{T}_\text{M} \cdot \mathbf{T}_\text{M})}},
\end{equation}
where the $\times$ denotes the multiplication between corresponding elements of two vectors, the $\cdot$ is the inner product of two vectors, and $\mathbf{1}$ indicates all-one vector.

Formally, we define the whole loss function of the proposed network as follows:

\begin{equation}
\label{eq1}
\begin{aligned}
L =\lambda_1 L_\text{cls}+\lambda_2 {L}_\text{dice}+\lambda_3 {L}_\text{atn},\\
\emph{s.t.} \; \sum_{i=1}^{3}\lambda _i=1 \; and \; \lambda_i \geq 0
\end{aligned}
\end{equation}
where ${L}_\text{cls}$ denotes the Binary Cross-Entropy loss (BCE-loss), and ${L}_\text{dice}$ is the Dice loss \cite{milletari2016v}, respectively. Specifically, the former two terms are defined as follows:
\begin{equation}\label{eq:cross-entr}
{L}_\text{cls}=-\frac{1}{N}\sum _{i=1}^{N}\Big( g_i\text{log}(p_i)+(1-g_i)\text{log}(1-p_i)\Big),
\end{equation}
\begin{equation}
{L}_\text{dice}=1-\frac{2\sum_{i=1}^{N}p_ig_i}{\sum_{i=1}^{N}(p_i+g_i)+\varepsilon},
\end{equation}
where $N$ is the total number of pixels in image. $p_i$ denotes the prediction of the $i$-th pixel in Crosslink-Net ($p_i=1$ indicates the target while $p_i=0$ indicates non-target). $g_i$ is the value of the $i$-th pixel in ground truth. $\varepsilon$ is a smooth factor.

\subsection{Discussion}
\label{subsec:ana}
To fully illustrate why our method works well, we now discuss on the benefits of double-branch encoder. The merit of our method includes three aspects.
\begin{itemize}
    \item Utilizing the vertical and horizontal kernels in two different branches can be expected to capture complementary information. Thus, Thus, in case the shape of targets can not be well observed from one single direction, the other direction might make up for it. Combining with the attention loss, the model learns in complementary directions and the learned information is simultaneously forced to correlate to the ground truth.
    \item Using two independent branches can better focus on vertical or horizontal information. Reversely, if we put the vertical and horizontal kernels into one encoder, the information will be merged after being passed through the first layer in the network, thus the information in the following features will become mixed. Besides, the double-branch encoder is also easier to optimize than its mixed single-branch counterpart.
\end{itemize}

To validate above analysis, we will replace the crosslink-kernels with square kernels in our Crosslink-Net in the following experiments. Moreover, we will also use the vertical and horizontal branches separately to construct two single-pathway encoder-decoder architectures, and then compare them with Crosslink-Net. In addition, to further highlight the advantage of our architecture, a ``fat'' single-pathway encoder-decoder architecture will also be constructed with the employment of both crosslink-kernel and some square kernels. Furthermore, we will also verify the complementation of two directions. More details are presented in the Section \ref{subsub:arc}.

\section{Experiments}
\label{sec:experiments}
To evaluate our model, we introduce four datasets with different targets in various shapes and sizes. The first three datasets are open-access ones, namely, cardiac MR dataset \cite{andreopoulos2008efficient}, kidney CT dataset Kits19 \cite{Heller2019the} and COVID-19 CT dataset \textbf{COVID19-I} \footnote{http://medicalsegmentation.com/covid19/}. The forth dataset is a private COVID-19 CT dataset, named as \textbf{COVID19-II}.

We first verify the efficacy of our attention loss and the advantages of our double-branch encoder and crosslink-kernels on the cardiac dataset. On this dataset, the heart tissue occupies a very small fraction of the entire MR image, and most boundaries of the myocardium (MYO), especially epicardium, are blurry, posing a great challenge for the segmentation model. Then, we evaluate the generalization and performance of Crosslink-Net on the small dataset COVID19-I that consists of 100 images. Finally, we compare Crosslink-Net with other SOTA methods on COVID19-II and Kits19 that are characterized by diverse shapes, various sizes, different positions and blurry boundaries.

\subsection{Datasets}
\textbf{COVID19-I.} This dataset is composed of 100 axial CT images (JPG format) with 99 of them being carefully annotated. The images on this dataset are small-size while diverse enough to train our deep model with merely some simple augmentations. The resolution of these images is $512 \times 512$ pixels and with three annotation labels: ground-glass, consolidation and pleural effusion. Due to the small size of this dataset, we follow Fan \etal \cite{fan2020inf} to take all three labels as one COVID-19 infection label.

\textbf{COVID19-II.} We also adopt another authentic COVID-19 CT volumes dataset independently collected from Shandong Chest Hospital, China. Each image is manually delineated by experienced radiologists. This dataset contains 96 contrast-enhanced abdominal CT volumes. The CT scan resolution is $512 \times 512 \times L$, where $L\in [2,26]$ is the number of slices along the long axis of body, and 5-26 slices in each sequence show lung infections. We cast the values of human tissues in the lung window (\eg, the window width and window level are 1500 HU and -500 HU) to the range of $[0,255]$ to obtain the JPG format.

\textbf{Kits19.} This is a CT dataset of MICCAI 2019 Kidney Tumor Segmentation Challenge \cite{Heller2019the}. There are total 210 cases with segmentation masks in the NIFTI format, each sample containing at least one kidney with tumor. We truncate the CT radio-density values to the range of [-160,240] HU to exclude irrelevant organs and objects, and save the CT slice in JPG format for our model.

\begin{figure}[ht]
 \centering
 \setlength{\abovecaptionskip}{-0.05cm}
  \includegraphics[width=0.47\textwidth]{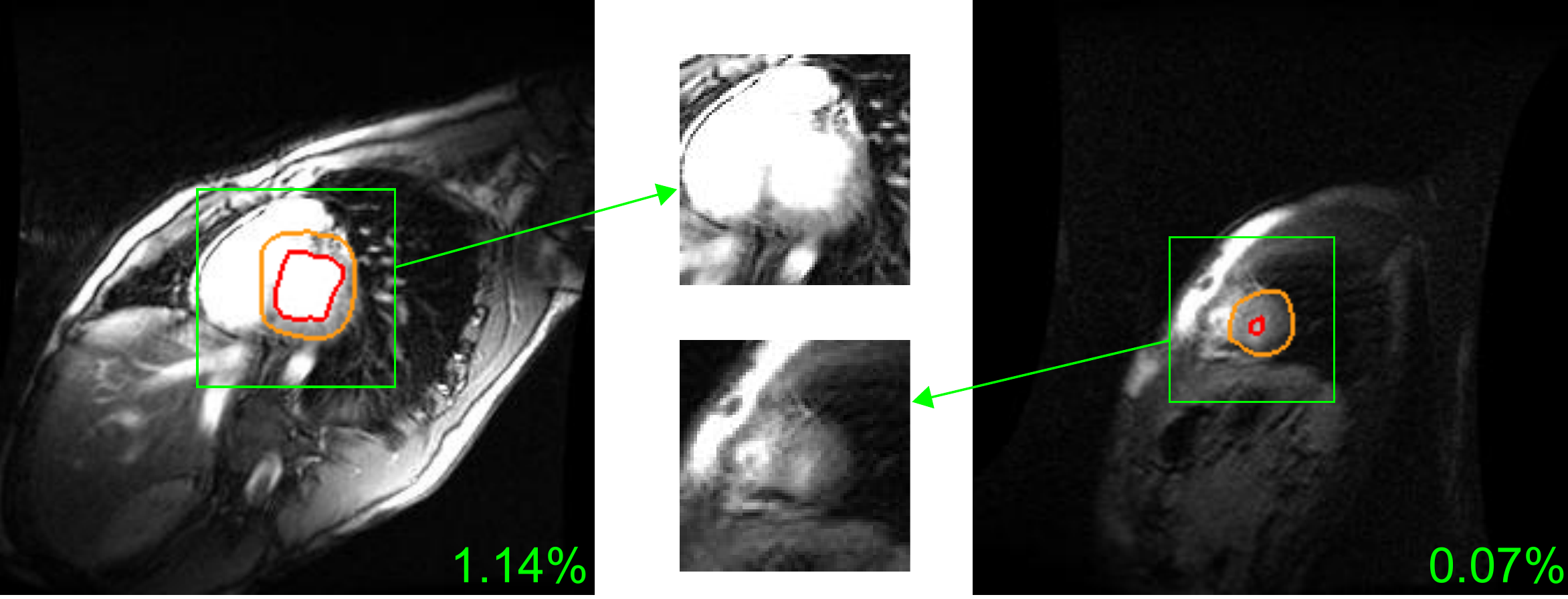}
 \caption{Examples of cardiac MR images. The red and orange contours are ground truth of endocardium and epicardium, respectively. The number in green is area fraction that LV cave occupies.}
 \label{fig:hart_exm}
\vspace{-0.3cm}
\end{figure}

\textbf{Cardiac MR Dataset.} This dataset is available from \cite{andreopoulos2008efficient} and comprised of cardiac MRI sequences for 33 subjects with a total of 7,980 MR slices, while the left ventricle (LV) is annotated in 4859 slices. The image resolution is $256 \times 256$ with the in-plane pixel size as 0.9-1.6 mm$^2$, and the inter-slice thickness as 6-13 mm. In each image, endocardial and epicardial contours are provided as the ground truth. In this dataset, the max area percentage of LV cave is no more than $2\%$ and the min less than $0.10\%$. Moreover, although different subjects do not show significant shape variations, most boundaries of MYO are unclear, as illustrated by examples in Fig. \ref{fig:hart_exm}.

\subsection{Setting}
\textbf{Implementation Details.}
For all datasets, the three-fold cross-validation is used to provide an unbiased estimation by randomly separating the dataset into training data and test data. For the two COVID19 datasets, we augment the training data through zooming images with a scale between $[0.5, 1.75]$, rotating between $[-30,30]$ degree and flipping horizontally. For the zoomed images, zero padding and random cropping are performed on small and large images, respectively. For the augmentation of cardiac dataset, we first enlarge the images twice and then crop them to $256 \times 256$, and then rotate them by 8 and -8 degrees.

The proposed Crosslink-Net is implemented with PyTorch on a NVIDIA Tesla P100 12GB Passive GPU with batch size as 2. The learning rate and weight decay is set as 0.5e-5 and 0.005, respectively, and Adam is adopted as the optimizer. The model weights are first initialized with Gaussian distribution and updated by employing the stochastic gradient descent algorithm during the training procedure.

\textbf{Metrics.} To evaluate the performance, six popularly-used metrics are calculated, namely, Dice similarity coefficient (DSC), over-segmentation rate (OR), under-segmentation rate (UR), Sensitivity (Sen.), Specificity (Spe.) and Hausdorff distance (HD) in millimeters (\ie, $mm$). For DSC, Sen. and Spe., the larger the better. For HD, OR and UR, the smaller the better. We report the first five metrics in the form of percentage (\ie, $\%$).

\subsection{Evaluation on Cardiac MR Dataset}
\label{sub:eva}

\subsubsection{Ablation Study}
We first evaluate our loss terms under different contributions to corroborate the necessity of every component in loss function (Eq. (\ref{eq1})). Our loss function contains three terms \ie, BCE-loss ${L}_\text{cls}$, Dice loss ${L}_\text{dice}$, and the proposed attention loss ${L}_\text{atn}$. We conduct an ablation study of these loss terms on the cardiac dataset, and specifically, set values for each $\lambda$ (\ie, $\lambda_1, \lambda_2, \lambda_3$) from a grid within $[0, 1]$ (0 to 1 in steps of 0.1). Some representative results are reported in Table \ref{table_loss}.

From the table, the following observations can be achieved: 1) Compared with using BCE-loss alone, the performance of the combination with attention loss or Dice loss is improved while that of the combination with attention loss is more desirable; 2) Compared with the performance of $\lambda_1 = 0.5$ and $\lambda_3 = 0.5$, that of the $\lambda_1 = 0.2$ and $\lambda_3 = 0.8$ is not improved obviously, indicating that Dice loss is still needed; 3) When the weight of the attention loss is fixed, a small portion of the Dice loss will be helpful; 4) $\lambda_1 = 0.4, \lambda_2 = 0.1$, and $\lambda_3 = 0.5$ yield the best performance, and hence the following experiments about Crosslink-Net on other datasets are all implemented with these weights.

\begin{table}[ht]
\setlength{\abovecaptionskip}{-0.2cm}
\setlength{\belowcaptionskip}{-0.5cm}
\caption{Metrics (\%) of Crosslink-Net on epicardial segmentation with different loss items. Note that larger DSC indicates better performance, and others are reverse.}
\label{table_loss}
\footnotesize
\begin{center}
\renewcommand\arraystretch{1.0}
\begin{tabular}{ccc|ccccc}
\toprule
$\lambda_1$&$\lambda_2$&$\lambda_3$&DSC&Sen.&Spe.&OR&UR\\
\midrule
1&0&0&93.75 &95.55&95.34&4.19&8.59 \\
0.5&0.5&0&93.88&95.52&95.38&4.32&6.47\\
0.2&0.8&0&93.92&95.76&95.65&3.99&7.51\\
0.5&0&0.5&94.41&95.92&95.71&\textbf{3.90}&6.60 \\
0.2&0&0.8&94.49&95.72&95.87&5.10&\textbf{5.17}\\
0.1&0.4&0.5&93.85&94.28&93.91&5.45&6.33\\
0.2&0.3&0.5&93.24&95.42&95.27&4.36&7.66\\
0.3&0.2&0.5&93.31&.94.59&94.34&5.12 &6.97\\
0.4&0.1&0.5&\textbf{95.28}&\textbf{96.58}&\textbf{96.41}&3.97&5.69\\
\bottomrule
\end{tabular}
\end{center}
\vspace{-0.3cm}
\end{table}

\begin{figure}[ht]
 \centering
 \setlength{\abovecaptionskip}{-0.1cm}
  \includegraphics[width=0.4\textwidth]{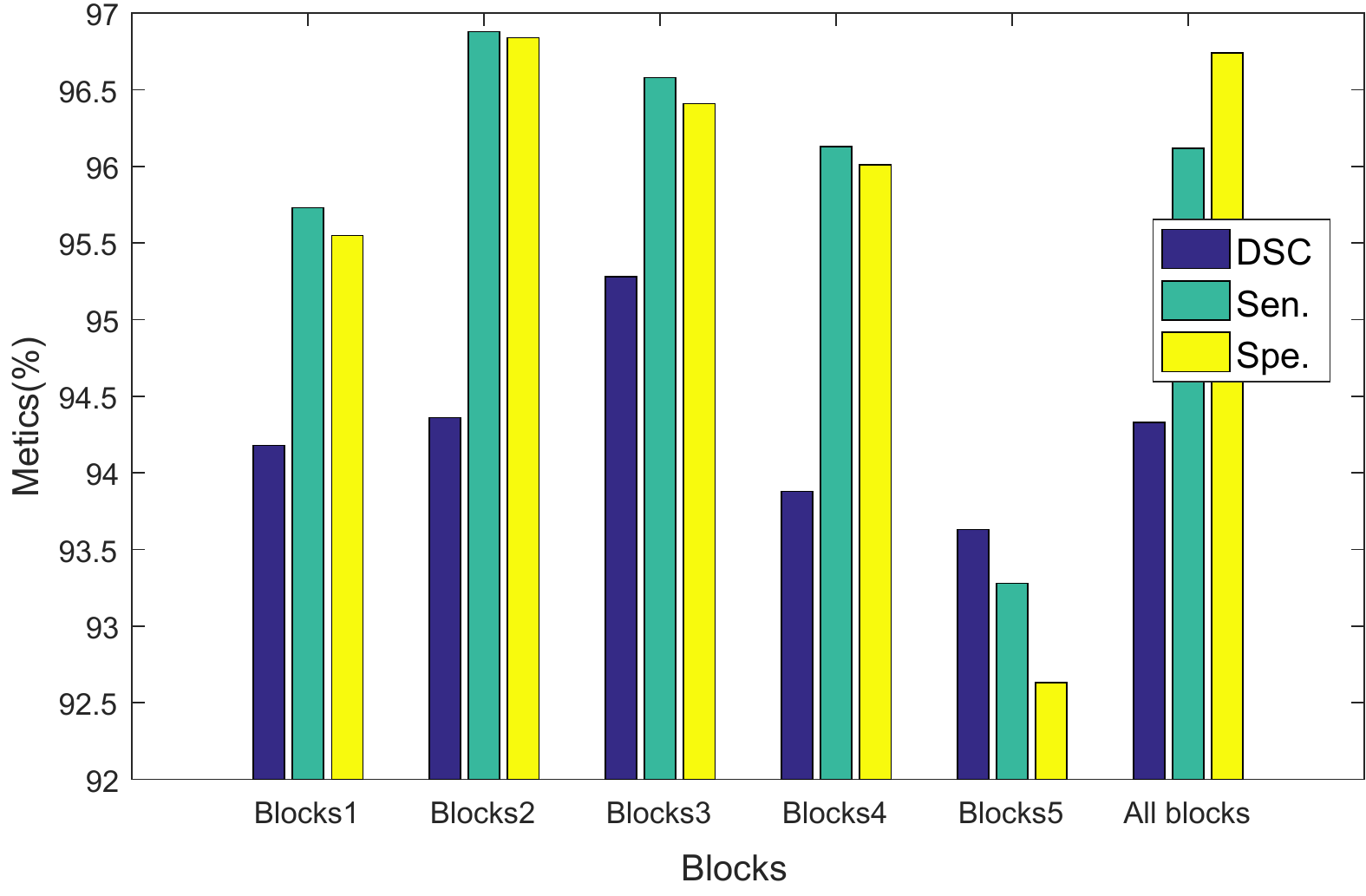}
 \caption{Metrics (\%) of Crosslink-Net on epicardial segmentation with attention loss being computed on different blocks.}
 \label{fig:atnblocks}
\vspace{-0.3cm}
\end{figure}

As stated in Section \ref{sub:loss}, the attention loss is computed on middle blocks (\ie, VCRB3 and HCRB3, denoted as \textbf{Blocks3} here), so here we verify it by evaluating the segmentation accuracy of Crosslink-Net with each block chosen from Blocks1 to Blocks5. We report the DSC, Spe., and Sen. of the model with the attention loss being computed on different blocks (Fig. \ref{fig:atnblocks}). As observed, although the change of each metric among blocks is not regular, metrics of blocks2 and blocks3 are relatively better than those of other blocks, while the DSC of blocks3 is more appealing. We also compute attention loss based on all five blocks, yet, the results are not superior to that of blocks3.

\subsubsection{Advantage of our Architecture}
\label{subsub:arc}

We first evaluate the advantage of crosslink-kernels in the double-branch encoder. Specifically, we adopt $3 \times 3$, $5 \times 5$ and $1 \times 1$ kernels in each convolution block of Crosslink-Net and keep other parameters unchanged, thus developing a new model called \textbf{SquareCrosslink}. Then, we compare our model with four single-branch encoder models: 1) Crosslink-Net containing only the vertical branch (named \textbf{VerCrosslink}), 2) Crosslink-Net containing only the horizontal branch (named \textbf{HorCrosslink}), 3) Crosslink-Net with a single-branch encoder containing all kinds of kernels (\ie, crosslink-kernels, $3 \times 3$, and $1 \times 1$), called \textbf{Double2SingleNet}, 4) \textbf{U-Net} \cite{ronneberger2015u}. For the sake of fairness, we implement all models with the BCE-loss term only. Meanwhile, we increase the number of the blocks of U-Net from four to five, each block composed of three continuous convolutions and residual connections. All models are implemented on epicardial segmentation. In addition, we also record the storage size of each model.

Table \ref{table_frame} reports the optimal performance of each model. The following observations can be obtained. 1) Crosslink-Net significantly outperforms SquareCrosslink . Although SquareCrosslink also employs double branches in its encoder, under the square kernels one branch fails to provide complementary information for the other, furthering confirming the advantage of crosslink-kernels. Nevertheless, compared with the single-encoder square model, i.e., U-Net, SquareCrosslink still shows the superiority of the double branches. 2) The performance of both VerCrosslink and HorCrosslink are inferior to Crosslink-Net yet competitive with Double2SingleNet, indicating that mixing information from different directions into one single block might bring confusion to the model, and hence a double-branch encoder is necessary. 3) Though also employing a crosslink-kernel, Double2SingleNet is poorer than Crosslink-Net in terms of performance. We believe that the overly “fat'' body of each block might have a negative impact during the gradient back-propagation, further demonstrating the necessity of a double-branch encoder. 4) Compared with Double2SingleNet and SquareCrosslink (the former has the same type of kernels with Crosslink-Net while the latter also has double encoders), our model has smaller storage size.

\begin{table}[ht]
\setlength{\abovecaptionskip}{-0.2cm}
\setlength{\belowcaptionskip}{-0.5cm}
\caption{The comparison among different architectures on epicardial segmentation with metrics (\%) being reported.}
\label{table_frame}
\footnotesize
\begin{center}
\renewcommand\arraystretch{1.0}
\begin{tabular}{c|cccccc}
\toprule
&DSC&Sen.&Spe.&OR&UR&Size\\
\midrule
VerCrosslink&92.71&95.24&96.24&\textbf{2.57}&9.97&60M\\
HorCrosslink&92.08&95.16&\textbf{95.77}&2.99&10.35&60M\\
U-Net \cite{ronneberger2015u}&91.21&94.00&94.18&4.97&10.02&135M\\
Double2SingleNet&92.68&95.47&95.27&4.28&9.27&178M\\
SquareCrosslink&92.97&95.08&94.70&5.73&\textbf{8.39}&165M\\
Ours&\textbf{93.75} &\textbf{95.55}&95.34&4.19&8.59&121M\\
\bottomrule
\end{tabular}
\end{center}
\vspace{-0.3cm}
\end{table}

\begin{figure}[ht]
 \centering
 \setlength{\abovecaptionskip}{-0.1cm}
  \includegraphics[width=0.47\textwidth]{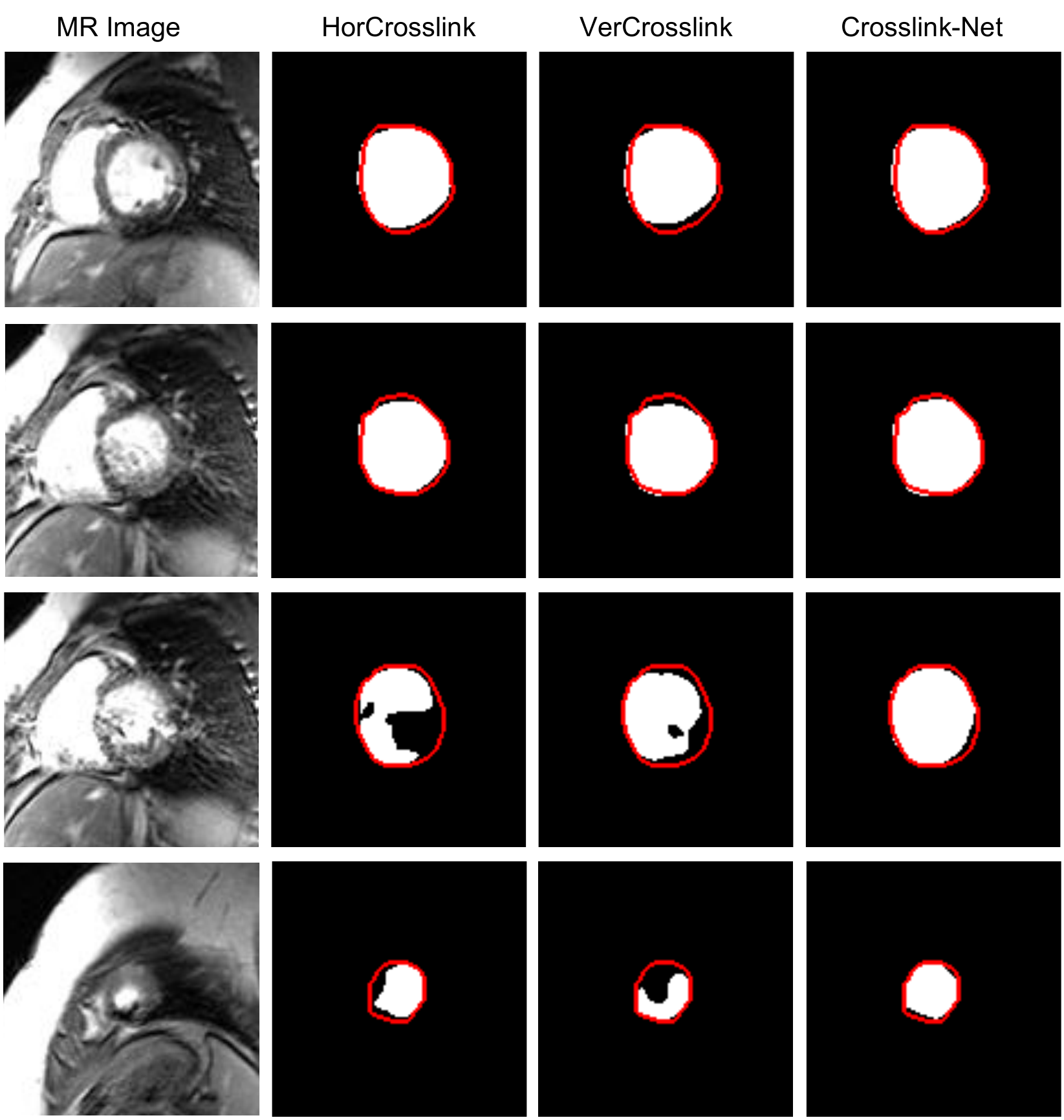}
 \caption{Some visual segmentation results of HorCrosslink, VerCrosslink and Crosslink-Net on MR cardiac dataset.The red contours are the ground truth.}
 \label{fig:hv}
\vspace{-0.3cm}
\end{figure}

We also verify the complementation between two branches of Crosslink-Net visually. Some visual results of Crosslink-Net with horizontal branch (HorCrosslink), double-branch and vertical branch (VerCrosslink) respectively on subject 24 of the cardiac dataset are illustrated in Fig. \ref{fig:hv}. The figure demonstrates that 1) most predictions of Crosslink-Net fit the ground truth better than those of HorCrosslink and VerCrosslink; 2) in case the shape information are not learned well from one direction, it can be learned from the other direction (\ie, the 2-nd and 7-th images); 3) learning from cross-direction through double-branch is better than simply combining the results of HorCrosslink and VerCrosslink.

\subsubsection{Comparison with Other Methods on Cardiac Dataset}

In this experiment, the LV cave and MYO are segmented. MYO is determined by endocardia and epicardium together, while LV cave is determined by the endocardia. On this dataset, we consider three SOTA models to compare with, namely, DenseUnet \cite{kaku2019DARTS}, GridNet-core \cite{zotti2019convolutional} and SPNet \cite{hou2020strip}. DenseUnet is a popular encoder-decoder model; GridNet is specially designed for cardiac MR image segmentation; and SPNet proposes a strip pooling mechanism and also employs non-square kernels.

We report the DSC, segmentation rate and distance metrics in Table \ref{tab:cardiac_com}. We observe that Crosslink-Net outperforms other models in terms of both LV cave and MYO segmentation. Meanwhile, we also observe that metrics of MYO are superior to that of LV cave, indicating that segmenting endocardia is a relatively hard task, compared with segmenting epicardium. This is probably due to the fact that segmentation models tend to ignore the small fraction of endocardia, whereas the performance of our model is much more desirable. In addition, SPNet \cite{hou2020strip} does not achieve satisfactory results, though it also adopts non-square kernels are. Strip pooling in SPNet performs well for long-range banded structure in scene parsing, yet, this model collects the whole row or column information as the context of the target. Thus, when the target is very small, local details might be neglected.

\begin{table}
\setlength{\abovecaptionskip}{-0.1cm}
\setlength{\belowcaptionskip}{-0.5cm}
  \centering
  \footnotesize
  \caption{Evaluation results for cardiac segmentation. Percentage (\%) is reported for DSC, OR and UR, and $mm$ for HD.}
  \label{tab:cardiac_com}
  \renewcommand\arraystretch{1.0}
    \begin{tabular}{p{0.75in}<{\centering}|p{0.12in}p{0.13in}p{0.13in}p{0.25in}|p{0.12in}p{0.13in}p{0.13in}p{0.25in}}
    \toprule
    \multirow{2}{*}{Method}&
    \multicolumn{4}{c}{LV Cave}&\multicolumn{4}{|c}{LV MYO}\\
    \cmidrule(lr){2-5} \cmidrule(lr){6-9}
    &DSC&UR&OR & HD & DSC &UR&OR & HD\\
    \midrule
    DenseUnet\tiny{\cite{kaku2019DARTS}}&89.89&11.36&08.71&04.52&91.15&10.20&07.27&05.14\\
    SPNet\tiny{\cite{hou2020strip}}&87.18&13.20&08.04&05.23&89.14&11.11&09.07&06.89\\
    GridNet-core\tiny{\cite{zotti2019convolutional}}&90.13&09.51&10.24&09.38&91.07&09.55&07.62&09.75\\
    Ours&\textbf{92.61}&\textbf{09.18}&\textbf{05.34}&\textbf{04.28}&\textbf{93.60}&\textbf{08.06}&\textbf{04.67}&\textbf{03.90}\\
    \bottomrule
    \end{tabular}
\vspace{-0.3cm}
\end{table}

\begin{figure}[ht]
 \centering
 \setlength{\abovecaptionskip}{-0.1cm}
  \includegraphics[width=0.4\textwidth]{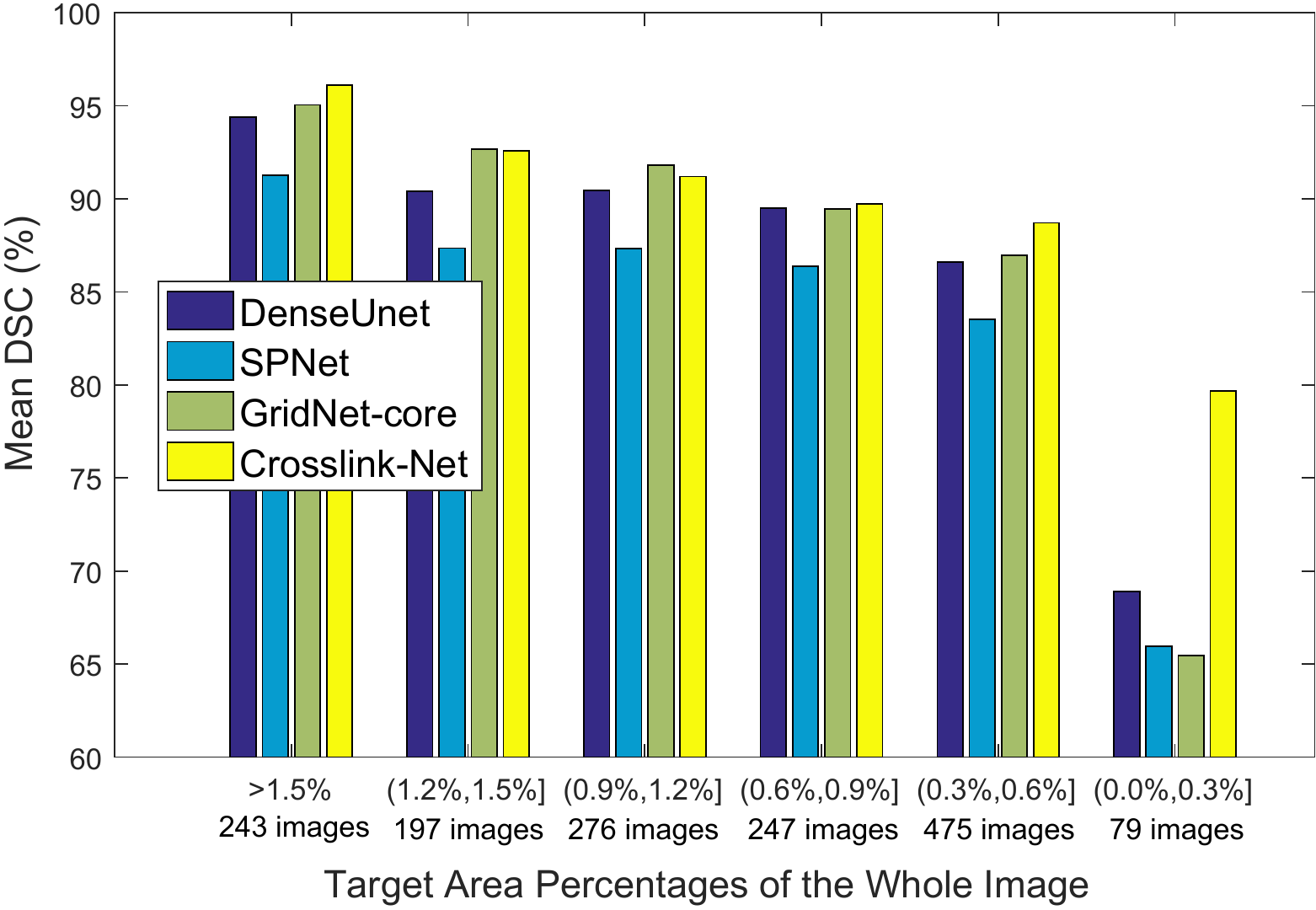}
 \caption{The comparison of mean DSC on LV caves with different area percentages. There are 1270 images in total.}
 \label{fig:hist}
\vspace{-0.3cm}
\end{figure}

Considering that LV cave is a small target, it is necessary to further verify the superiority of Crosslink-Net for the LV cave segmentation. Therefore, we randomly select several subjects from one cross-validation experiment to observe the distribution of DSC. The test set contains subject 24-32, with a total of 1270 images. As illustrated in Fig. \ref {fig:hist}, we list the target area percentages versus mean DSC values of different models. It is obvious that Crosslink-Net shows remarkable superiority especially on targets which take area fractions less than $0.6\%$, further confirming the advantage on small target segmentation.

\begin{figure}[ht]
 \centering
 \setlength{\abovecaptionskip}{-0.1cm}
  \includegraphics[width=0.48\textwidth]{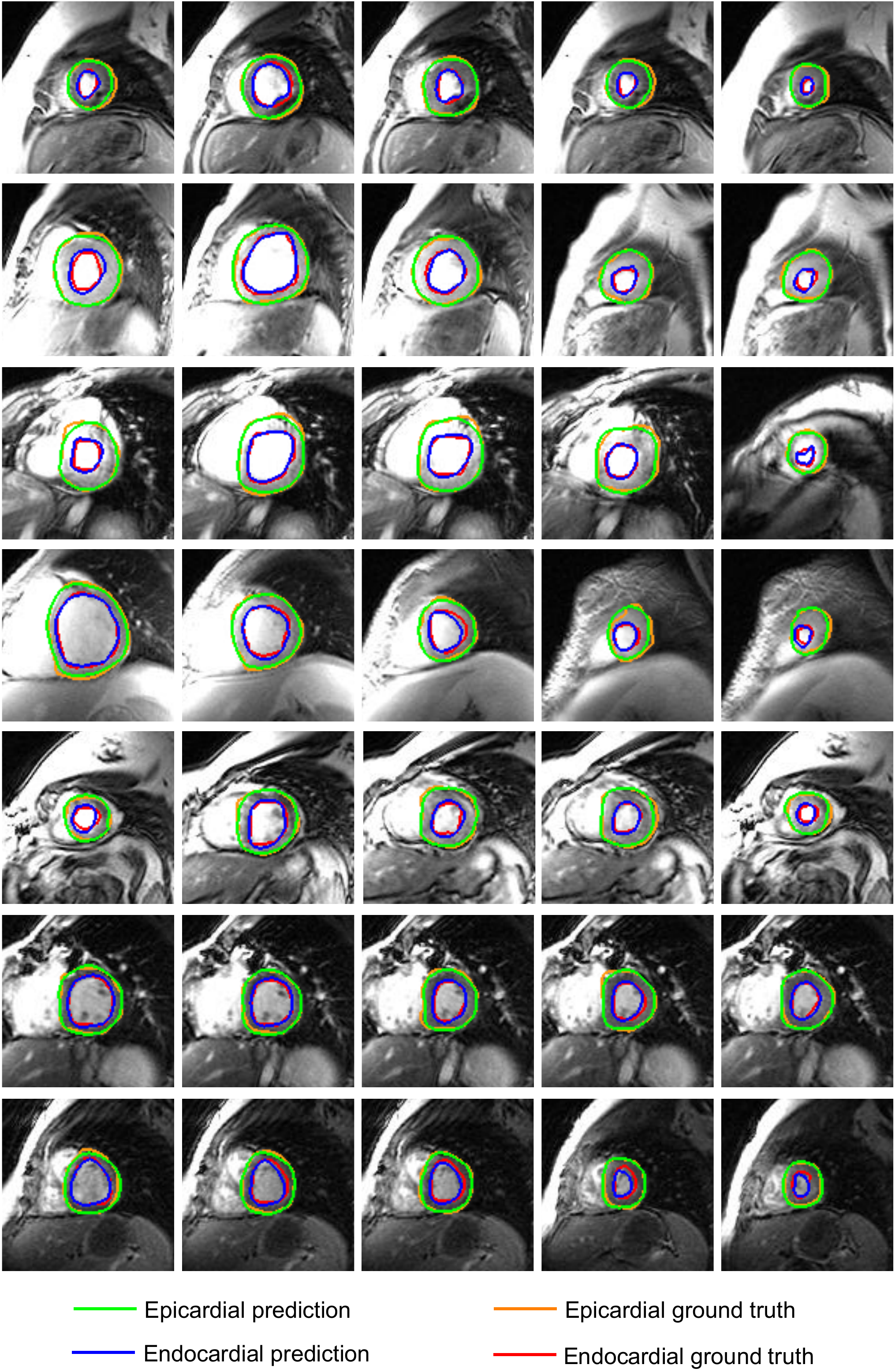}
 \caption{Some predictions of Crosslink-Net on cardiac dataset.}
 \label{fig:cardiac}
\vspace{-0.3cm}
\end{figure}

We also present some visual results of our models as shown in Fig. \ref{fig:cardiac} with several representative samples from each sequence of subject 24-30. As consistent with Fig. \ref{fig:hist} and Table \ref{tab:cardiac_com}, the counterparts depicted by our model could fit ground truth well.

\subsection{Comparison with Other Methods on COVID19-I Dataset}

To further evaluate the effectiveness and generalization of our model on small datasets, we have compared the Crosslink-Net with three models, \ie, U-Net \cite{ronneberger2015u}, Inf-Net \cite{fan2020inf} and Crossover-Net \cite{yu2020crossover}, on COVID19-I. The first is a classical encoder-decoder model, which we have modified in the same way as we did in Section IV-C2. The second is a SOTA model specially designed for COVID-19 CT images segmentation task, with a pre-trained model (e.g., ResNet50 and VGGNet16) as its backbone. The code of Inf-Net is provided by its original manuscript and we modify the augmentation of training data in the source code. The third is a patch-based model with patches instead of images as its input.

The results of all these models are illustrated in Table \ref{table_COVI}. We can observe that Crosslink-Net achieves the best Dice and under-segmentation rate. This result demonstrates that, compared with both the patch-based model Crossover-Net and the pre-trained model Inf-Net, the small dataset has not degraded the performance of our model. The high DSC and Sen. indicate that Crosslink-Net is not prone to mis-segmenting. However, the unremarkable Spe. shows that, in small dataset, more efforts should be made for our model to learn fufficient information so as to distinguish the infected areas from other tissues.

\begin{table}
\setlength{\abovecaptionskip}{-0.2cm}
\setlength{\belowcaptionskip}{-0.5cm}
\caption{The comparison among different methods on COVID19-I (percentage (\%) is reported).}
\label{table_COVI}
\footnotesize
\begin{center}
\renewcommand\arraystretch{1.0}
\begin{tabular}{c|ccccc}
\toprule
Method&DSC&Sen.&Spe.&OR&UR \\
\midrule
U-Net \cite{ronneberger2015u}&62.09&72.75&72.17&18.81&34.66\\
Crossover-Net\cite{yu2020crossover}&72.80&78.35&\textbf{78.20}&\textbf{14.22}&24.33\\
Inf-Net \cite{fan2020inf}&70.17&75.83&73.38&18.73&25.82\\
Ours&\textbf{73.40}&\textbf{79.17}&77.79&14.37&\textbf{20.13}\\
\bottomrule
\end{tabular}
\end{center}
\vspace{-0.3cm}
\end{table}

\begin{figure*}
 \centering
  \includegraphics[width=0.95\textwidth]{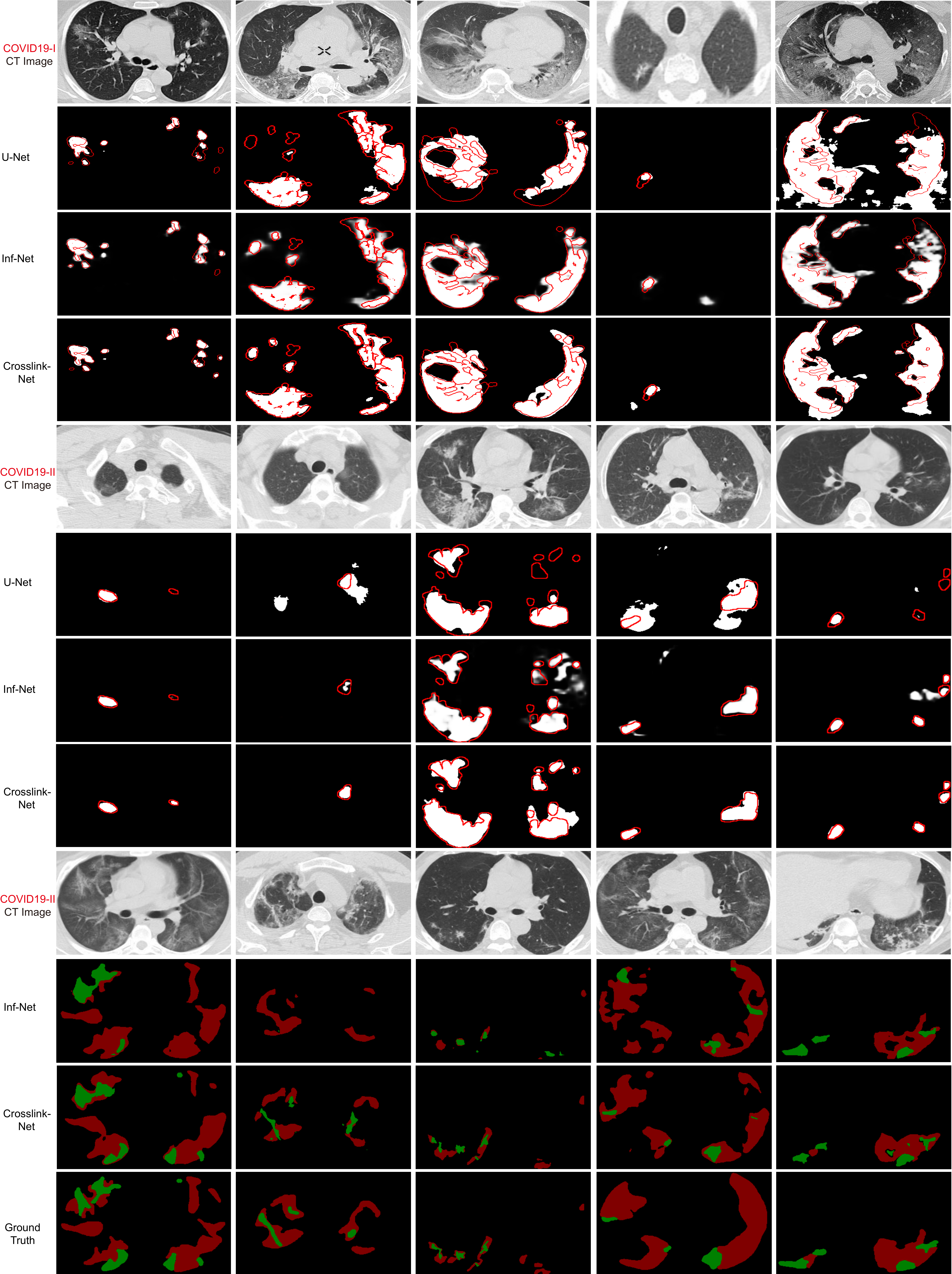}
 \caption{Some qualitative results of different methods on COVID19-I and COVID19-II. The red contours are ground truth of lung infections while the red and green regions in the last three rows indicate the ground-glass and consolidation, respectively.}
 \label{fig:COV}
\end{figure*}

We also visualize some qualitative results in Fig. \ref{fig:COV}. These results are consistent with the quantitative results in Table \ref{table_COVI}. For example, most of the prediction results of Crosslink-Net are close to the ground truth, which is consistent with the relatively high Dice. Besides, some normal tissues are predicted as the infect areas occasionally, which might lead to a lower Specificity. Obviously, some small infected areas remain mis-segmented, mainly due to insufficient training data.

\subsection{Comparison with Other Methods on COVID19-II Dataset}

Since COVID19-II is much larger than COVID19-I, we conduct the multi-target segmentation on this dataset. As the results of each model listed in Table \ref{table_COV_met} have shown, Crosslink-Net outperforms both U-Net and Inf-Net in terms of both DSC and Spec. by a large margin, especially in terms of the DSC of ground-glass segmentation. Considering that boundaries of ground-glass are blurry, we attribute this improvement to our crosslink-kernels and complemented double branches for their abilities of capturing detailed information. Meanwhile, although consolidation segmentation is more challenging compared to ground-glass, the proposed model is still competitive with Crossover-Net. The Crossover-Net, a patch-based model that converts the segmentation task into the classification of the central pixel of patches, is great for capturing the local information around the central pixel. This mechanism can help the model better focus on blurry boundaries, therefore, Crossover-Net performs very similar to Crosslink-Net on both COVID19-I and COVID19-II. However, this type of patch-based model usually suffers from redundant computations.

\begin{table*}[ht]
\setlength{\abovecaptionskip}{-0.05cm}
\setlength{\belowcaptionskip}{-0.5cm}
  \centering
  \footnotesize
  \caption{Prediction performance comparison on COVID19-II (percentage (\%) is reported).}
  \label{table_COV_met}
  \renewcommand\arraystretch{1.0}
    \begin{tabular}{c|c|ccc|ccc|ccc|ccc}
    \toprule
    \multirow{2}{*}{Method}&\multirow{2}{*}{Venue}&
    \multicolumn{3}{c}{Infection}&\multicolumn{3}{|c|}{Ground-glass}&\multicolumn{3}{c|}{Consolidation}&\multicolumn{3}{c}{Average}\\
    \cmidrule(lr){3-5} \cmidrule(lr){6-8} \cmidrule(lr){9-11} \cmidrule(lr){12-14}
    & &DSC &Sen. &Spec. &DSC &Sen. &Spec. &DSC &Sen. &Spec. &DSC &Sen. &Spec. \\
    \midrule
    U-Net \cite{ronneberger2015u}&MICCAI2015&62.05&74.52&65.51&53.87&52.64&58.22&38.51&47.28&46.12&51.47& 58.14&56.61\\
    Crossover-Net\cite{yu2020crossover}&PR2021&72.10&80.23&77.37&66.01&78.46&77.40&\textbf{61.78}&66.29&\textbf{49.33}&66.63&\textbf{74.99}&68.03\\
    Inf-Net\cite{fan2020inf}&TMI2020&70.03&\textbf{81.36}&76.49&63.63&70.76&66.04&58.52&\textbf{68.24}&36.76&64.72&73.12&59.76\\
    Ours&--&\textbf{72.50}&79.85&\textbf{79.79}&\textbf{66.69}&\textbf{79.00}&\textbf{77.56}&60.84&64.48&48.15&\textbf{67.01}&68.52&\textbf{68.50}\\
    \bottomrule
    \end{tabular}
\vspace{-0.3cm}
\end{table*}

To further indicate the superiority of our model, we provide some qualitative results on lung infection, ground-glass and consolidation segmentations. As shown in the middle four rows of Fig. \ref{fig:COV}, for the lung infections, segmentation results of Crosslink-Net are close to the ground truth with much less mis-segmented tissues. In contrast, U-Net performs unsatisfactorily, with a large number of tissues being ignored. Whereas, though Inf-Net improves the results, its performance is still not desirable with more mis-segmented tissues. However, Inf-Net tends to label more irrelevant regions on ground-glass (the third image), probably because its DSC is inferior to ours while its Sen. superior to ours.
In terms of multi-class segmentation, our Crosslink-Net fits ground truth better than Inf-Net does, especially for the consolidation, as illustrated in the last three rows.

\subsection{Results on Kits19 dataset}
For the kidney tumor segmentation task, the SOTA models could be roughly classified into two types: the 3D end-to-end models and the 2D/3D coarse-to-fine models. Therefore, we compare with five 3D models with first three being end-to-end ones: The first is MSS-Unet \cite{Zhao2019Multi}, with a patch size of $192 \times 192 \times 48$ and using 3D U-Net; the second is a pre-activation residual 3D U-Net (PR3D-Unet) \cite{Isensee2019an} with a patch size of $160 \times 160 \times 80$; and the third is FCN-PPM \cite{yang2018auto}, which first abstracts the 3D ROI ($150 \times 150$) via other methods and then segments the kidney and tumor with deep model. The forth and fifth  are  3D coarse-to-fine models, \ie, CU-Net \cite{li2019fully} and C3D-Unet \cite{zhang2020cascaded}, both of which train a 3D U-Net to segment kidney as the ROI input for the second 3D U-Net for tumors. We follow the steps of the coarse-to-fine methods, setting $192 \times 192 \times 30$ both as the input patch for all five 3D models and as ROI for CU-Net and C3D-Unet, while setting $192 \times 192$ as ROI for our Crosslink-Net. For all the reference models, we implement them according to their original manuscripts.

We list the DSC of each model in Table \ref {tab:kits19}. As observed, for the tumors, although the 3D information is also utilized in the SOTA models, our model is competitive with or even slightly better than them. For the kidney segmentation, our model achieves a slight improvement upon the 3D models. It is also worth noting that in the coarse stage, the coarse image input to our model is $512 \times 512$ while it is $192 \times 192 \times 30$ for the C3D-Unet, hence, the tumor DSC of C3D-Unet is apparently higher than ours in the coarse stage. However, the DSC of our model on kidney segmentation in the coarse stage remains competitive with the final result of FCN-PPM, showing the advantage of our model again.

\begin{table}[ht]
\setlength{\abovecaptionskip}{-0.3cm}
\setlength{\belowcaptionskip}{-0.9cm}
\caption{DSC(\%) of different models on Kits19 dataset. Blue numbers are results of coarse segmentation.}
\center
\renewcommand\arraystretch{1.0}
\label{tab:kits19}
\footnotesize
\begin{tabular}{p{0.9in}<{\centering}|m{0.3in}m{0.3in}<{\centering}|m{0.3in}<{\centering}m{0.3in}<{\centering}}
\toprule
Method & \multicolumn{2}{c}{Kidney} & \multicolumn{2}{|c}{Tumor} \\
\midrule
MSS-Unet \cite{Zhao2019Multi} & \multicolumn{2}{c}{96.15}  & \multicolumn{2}{|c}{80.01} \\
FCN-PPM \cite{yang2018auto}& \multicolumn{2}{c}{92.87}      & \multicolumn{2}{|c}{79.22}     \\
PR3D-Unet \cite{Isensee2019an}& \multicolumn{2}{c}{96.98}      & \multicolumn{2}{|c}{84.05}     \\
\midrule
CU-Net \cite{li2019fully}&\re{88.76}  &96.88&\re{43.53}&80.62\\
C3D-Unet \cite{zhang2020cascaded}&\textbf{\re{96.82}}&97.16&\textbf{\re{72.10}}&83.25\\
Ours&\re{92.35} &\textbf{97.21}&\re{52.71}&\textbf{84.11}\\
\bottomrule
\end{tabular}
\vspace{-0.3cm}
\end{table}

\begin{figure}
 \centering
 \setlength{\abovecaptionskip}{-0.2cm}
  \includegraphics[width=0.47\textwidth]{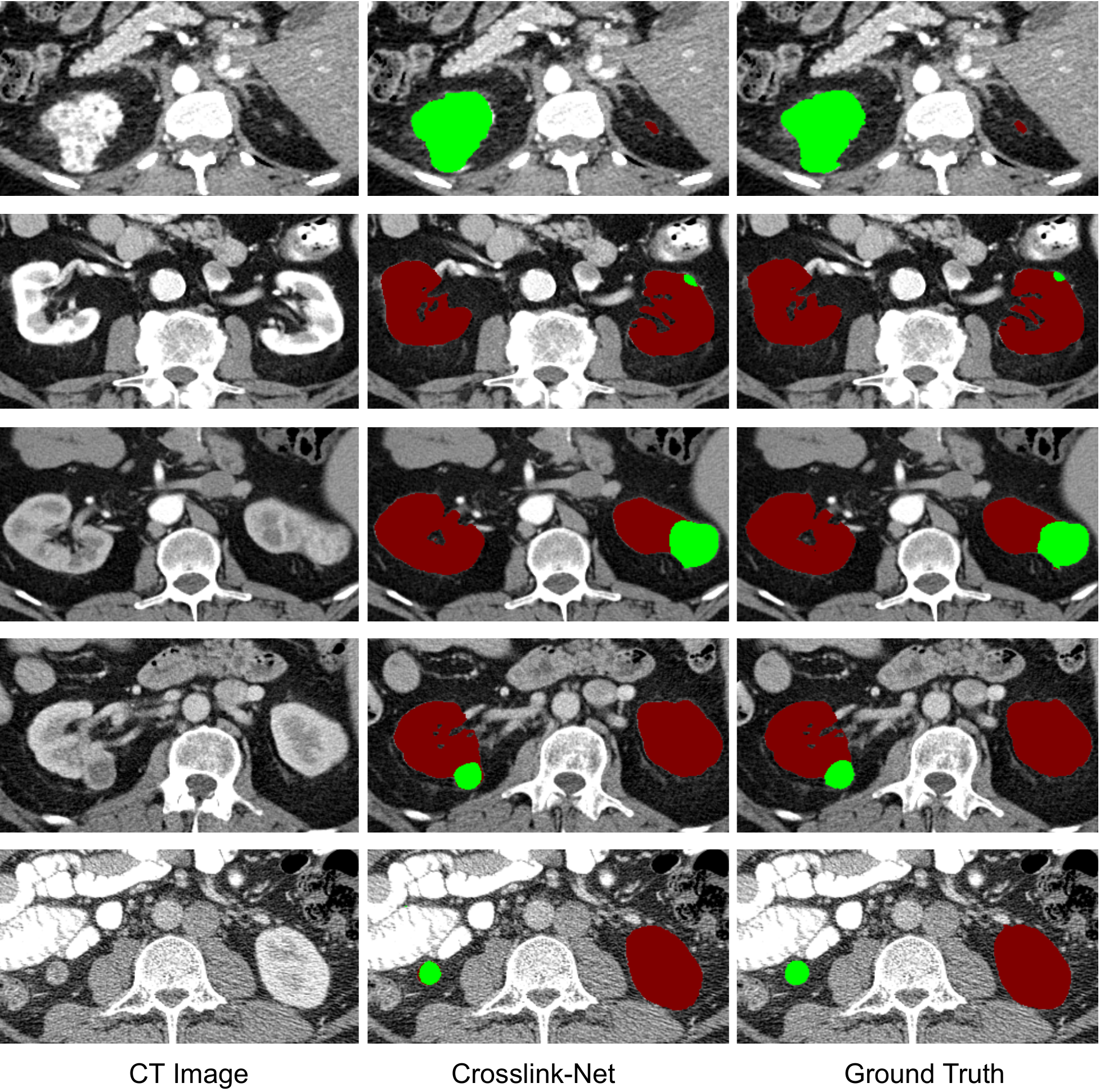}
 \caption{Comparison between ground truth and predictions of Crosslink-Net.The red and green regions indicate the kidney and tumors, respectively.}
 \label{fig:kits}
\vspace{-0.3cm}
\end{figure}

According to some representative examples of visual results shown in Fig. \ref{fig:kits}, we can obtain the following three observations: 1) First, our model can accurately segment even very small kidney region, which reveals our method can better handle small targets with our attention loss. 2) second, in the third and forth cases, although the tumor is strongly connected to the renal parenchyma with no boundaries, Crosslink-Net still shows desirable performance, indicating its great ability to capture detailed information. 3) In the fifth case, although the image is disturbed by noises, Crosslink-Net still can segment it well, showing the robustness of our model to noises.

\section{Conclusion}
\label{sec:conclusion}
In this paper, we propose a novel and efficient medical image segmentation framework with a double-branch encoder and a typical decoder, \ie, Crosslink-Net. In particular, the proposed double-branch encoder can simultaneously capture both the vertical and horizontal information with non-square vertical and horizontal kernels, so as to effectively deal with different shapes with blurry boundaries. To better handle the small target regions, we further develop an attention loss to enhance the correlation between vertical and horizontal branches and the ground truth in the encoding stage. Experiments on three public datasets and one private dataset show that our model performs well in cases of targets with various shapes and sizes. Moreover, we also validates the efficacy of the proposed crosslink-kernels, double-branch architecture, and the loss function.

 Considering that selecting an optimal hyper-parameter is difficult, our future work is to automatically learn the hyper-parameter of each term in loss function with the meta-learning scheme, rather than set them empirically during the training process.
%
%
%

\ifCLASSOPTIONcaptionsoff
  \newpage
\fi



%
%
\bibliographystyle{IEEEtran}
\bibliography{egpaper_final}

\end{document}